\documentclass{eslart}
\usepackage{psfig}
\usepackage{amssymb}

\begin{document}
\def\D0{D\O}                        
\newenvironment{symbolfoot}
    {\renewcommand{\thefootnote}{\fnsymbol{footnote}}}

\begin{frontmatter}
\title{Time Resolution and Linearity Measurements for a Scintillating 
Fiber Detector Instrumented with VLPC's}

\author[Fermilab]{A. Bross}
\author[CBPF]{A. Chaves}
\author[CBPF]{J. Costa}
\author[Fermilab]{M. Johnson}
\author[CBPF]{L. Moreira}
\author[Maryland]{J. Thompson}
\address[Fermilab]{Fermi National Accelerator Laboratory,
Batavia, IL 60510}
\address[CBPF]{LAFEX/CEFET - EN, Rio de Janeiro, Brazil}
\address[Maryland]{Department of Physics, University of Maryland,
College Park, MD 20742}

\begin{abstract}
The time resolution for charged particle detection is reported for 
a typical scintillating fiber detector instrumented with 
Rockwell HISTE-IV Visible Light Photon 
Counters.  The resolution measurements are shown to agree with a simple Monte 
Carlo model, and the model is used to make recommendations for improved 
performance.  In addition, the gain linearity of a sample of VLPC devices 
was measured.  The gain is shown to be linear for incident light 
intensities which produce up to approximately 600 
photoelectrons per event.
\end{abstract}
\end{frontmatter}

\section{Introduction}
Particle detectors consisting of scintillating fiber arrays instrumented with
Visible Light Photon Counters (VLPC's) are being constructed for the 
next generation of collider experiments \cite{Ruchti96}. 
The fast signal response of these detectors 
allow their use in environments with high event rates.  Furthermore, 
the low noise and linear gain characteristics of the VLPC's
allow accurate photon counting.  This report presents
measurements of the time resolution for charged particle detection for a 
practical scintillating fiber detector
instrumented with VLPC's and makes predictions of expected future 
performance.  This information is critical to determine the efficacy of using 
time difference measurements on such systems to calculate the longitudinal
position of a charged particle or shower impacting the fiber.  
Also reported are linearity measurements of the VLPC response for light 
intensities up to $1 \times 10^{5}$ incident photons per event. This linearity
study is relevant to detectors using VLPC's for calorimetry.

The VLPC is a solid-state photomultiplier \cite{Petroff87} 
optimized for visible
light detection and developed by 
Rockwell International \cite{Rockwell}.  These 
devices provide excellent 
performance as light detectors since they have a fast gain of approximately 
$1.2 \times 10^{4}$ per photoelectron (within 40~ns) with a quantum efficiency 
of up to 80\%~\cite{model}.  Their major
drawback as photodetectors is the requirement that they be operated at liquid 
helium temperatures.

The measurements reported here stem from engineering studies by a subgroup of
the  \D0 collaboration which plans to have two scintillator/VLPC systems for 
its upgraded detector: a scintillating fiber central tracker for charged 
particle tracking and triggering, and a 
preshower detector for electron identification and triggering.  Preliminary 
operating conditions for the proposed systems have been chosen based on
extensive studies of the VLPC \cite{Characterization}. 
The operating parameters chosen for these timing and linearity measurements 
reflect those plans.  The VLPC's are HISTE-IV models with 1 mm$^2$ pixels 
which were
mounted in a cassette \cite{CosmicRay} and located in a liquid helium cooled 
cryostat.  The cassette contained one 50~cm long clear multiclad
polystyrene fiber for each pixel which
provided a light path from a room temperature fiber connector.  The VLPC's were
operated at 6.5~K with a bias voltage of 6.5~V.  The light source was 
commercial polystyrene scintillator doped with 1\% p-terphenyl 
and 1500~ppm of 3-hydroxyflavone (3HF).   This combination
produced visible light peaked 
in wavelength at 530~nm \cite{SpectralResponse} with a fluorescence 
decay time of 8.2~$\pm$~0.2~ns \cite{DecayTime}; 
this wavelength is well matched to the spectral response of the 
VLPC.  

\section{Timing Test: Experimental Setup}

\begin{figure}
\centerline{\psfig{file=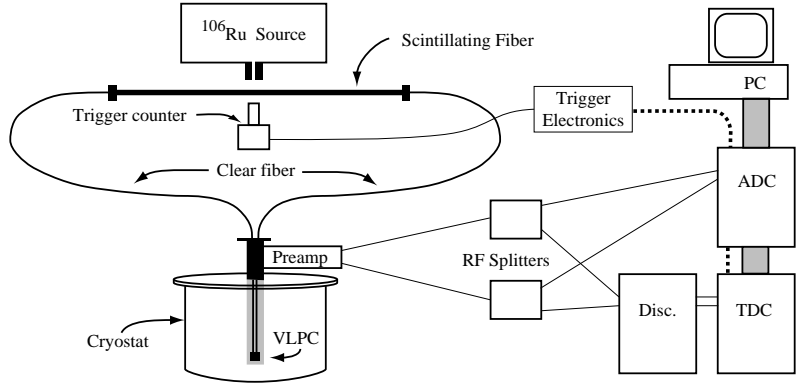,height=2.0in}}
\caption{
Block diagram of experimental setup for system time resolution studies.
\label{timset}}
\end{figure}

The basic experimental setup for the timing tests 
is shown in Fig.~\ref{timset}.  

The system timing tests required that scintillating fiber act as the light
source to account for light dispersion within the fiber.  The
fiber was commercial double-clad 3HF scintillating 
fiber~\cite{Multiclad} as proposed for the upgraded \D0 tracking system.  
The fiber was instrumented on
both ends, with the time difference ($\Delta t$) between receipt of the signal 
at each end being the quantity of interest.  Constraints within the 
existing \D0 detector require that the cryostats containing the VLPC's be 
physically separated from the tracking detector, so in some cases a clear 
fiber of 8~m length was used as a light guide between the scintillating fiber 
and the cassette containing the VLPC's.  Relevant parameters for the fibers 
are given in Table~\ref{table_fiber}.  Mineral oil was used as the optical 
coupling between diamond-polished fiber ends within mating connectors.

\begin{table}
\caption{Specifications for scintillating and clear fiber.}
\label{table_fiber}
\begin{tabular}{l c c c}
\hline
 & & Scintillating Fiber & Clear Fiber \\
\hline
Core & Diameter ($\mu$m) & 755 & 875 \\
& Principle composition & polystyrene & polystyrene \\
& Dopants & p-terphenyl 1\% & \\
& & 3HF 1500~ppm &\\
& Index of refraction & 1.59 & 1.59 \\ 
\hline
First Cladding & Thickness ($\mu$m) & 20 & 23 \\
& Composition & \multicolumn{2}{c}{Acrylic (PMMA)} \\
& Index of Refraction & \multicolumn{2}{c}{1.48}  \\ 
\hline
Second Cladding & Thickness ($\mu$m) & 20 & 23 \\
& Composition & \multicolumn{2}{c}{Fluorinated PMMA} \\
& Index of refraction & \multicolumn{2}{c}{1.42}  \\ 
 \hline
\end{tabular}
\end{table}

Light was generated within the scintillating fiber by illuminating it with
a collimated $^{106}$Ru radioactive source.  Ruthenium provides the highest 
energy beta rays naturally available and was preferred to cosmic ray muons 
because of the increased event rate.  A scintillator paddle instrumented 
with a photomultiplier tube and located in the fiber's shadow from the 
radioactive source served as the system trigger.

In spite of its high gain, the signal from the VLPC must be amplified, 
especially when used for low light intensities when only a few photoelectrons
are produced.
In such an application, the amplifier operates in pulse mode, meaning 
that the time integral of each burst of current, the charge, is converted to a
voltage value.  This needs to be done with a minimum of additional noise and 
with sufficient speed to minimize the effect on system time resolution.
Measurements of signal pulses directly out of the cassette for high 
light intensities yielded rise times of 3--5~ns. (Internal rise times within
the VLPC may be somewhat faster but the limiting time is that observed outside
the cassette.  The frequency response of the signal cable within 
the cassette was measured to drop by 2.5~dB at 100~MHz.) 
The 3~ns figure yields an equivalent bandwidth of 120~MHz
which sets a lower limit on the amplifier's performance.  
The final constraint on amplifier performance 
is the gain necessary to provide sufficient output voltage for the 
following stages of instrumentation and analysis.  Note that it is difficult 
to obtain acceptable signal-to-noise performance for
amplifiers with high gain, high sensitivity, and high bandwidth.

\begin{figure}
\centerline{\psfig{file=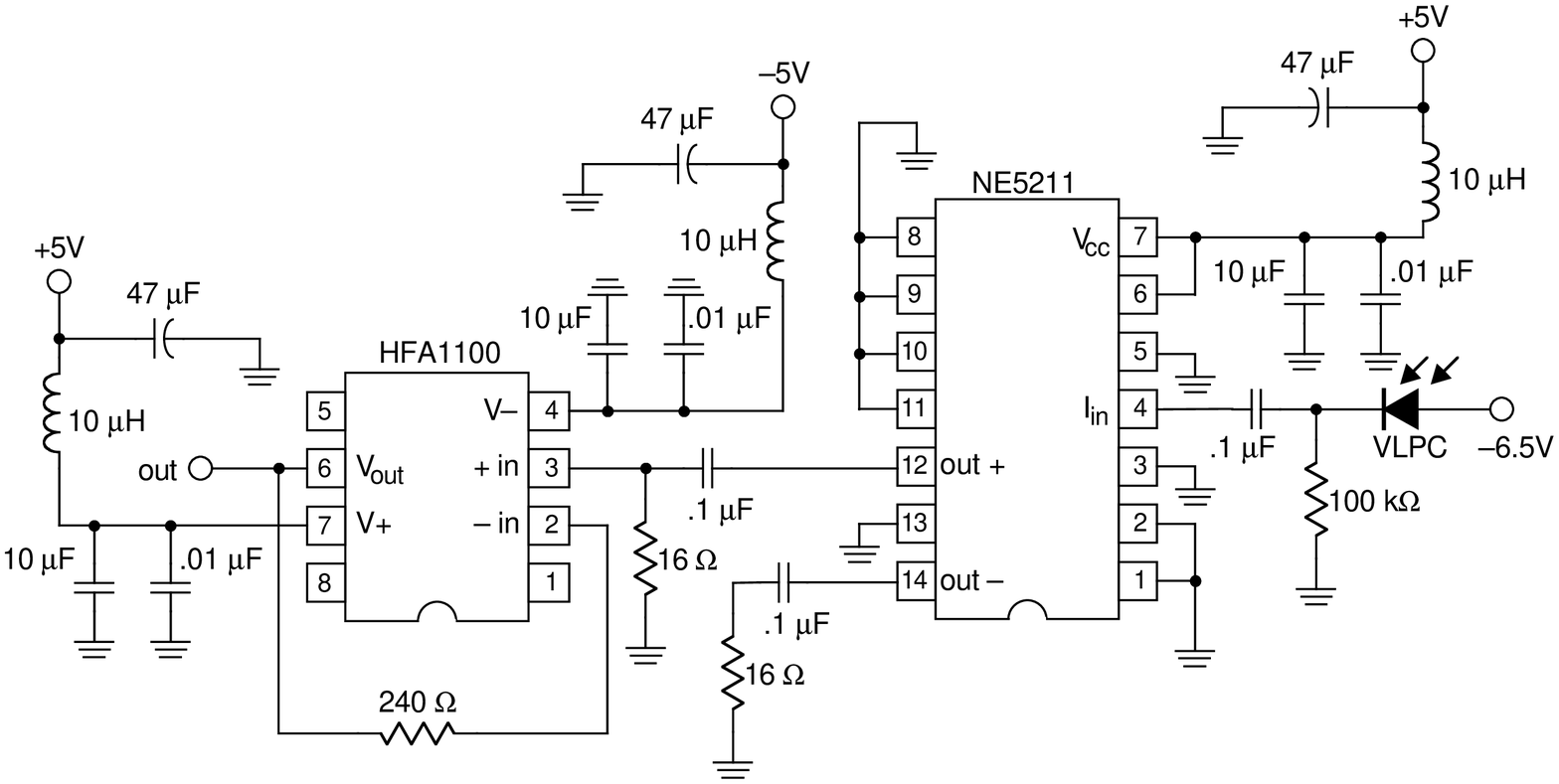,height=2.6in}}
\caption{
Schematic of preamplifier circuit used for system timing measurements.
\label{Phil_Harr}}
\end{figure}

Several amplifiers were tested, but no single device met all constraints.
The best results
were obtained with a combination of two commercial devices:
the Philips NE5211 \cite{Philips} transimpedance amplifier 
to convert the current pulse into a voltage pulse followed by 
the Harris HFA1100 \cite{Harris} current feedback
amplifier to amplify the voltage pulse.
The two stage circuit is shown in Fig.~\ref{Phil_Harr}.
The NE5211, as a single device, had better time resolution 
than the other options but also had relatively
low gain which was unacceptable for light intensities below ten detected
photons. However, the amplifier's low input noise combined with its
high sensitivity make it desirable as a first stage particle detector
preamplifier since the most important noise sources are at the detector
front-end.   
The primary purpose of the second stage was to provide voltage 
amplification with a minimum impact on the system time resolution. 
As a current feedback amplifier (or transimpedance amplifier) 
the HFA1100 is primarily intended for current-to-voltage conversion, but
for this test the device was configured as a non-inverting voltage amplifier. 
The main features of these devices are listed in Table~\ref{table_preamps}.

\begin{table}
\caption{Specifications for the commercial amplifiers.
The bandwidth of the HFA1100 is given as configured for the 
amplifier circuit used.}
\label{table_preamps}
\begin{tabular}{c c c}
\hline
& NE5211 & HFA1100 \\
\hline
Bandwidth & 180~MHz & 120~MHz \\
Input Noise & 1.8~$\mathrm{pA}/\sqrt{\mathrm{Hz}}$ & 
4~$\mathrm{nV}/\sqrt{\mathrm{Hz}}$ \\
Input Impedance & 200~$\Omega$ & 10~$\Omega$ (Input~--) \\
&& 50~K$\Omega$ (Input~+) \\
Input Capacitance & 4~pF & 2pF \\
Transimpedance Gain & $R_{T}= 28$ k$\Omega$ (differential output) & 
$R_{T}= 500$ k$\Omega$ (open loop)\\
& $R_{T}= 14$ k$\Omega$  (single ended output) & \\
\hline
\end{tabular}
\end{table}


The actual time resolution may be better than that measured because the VLPC
signals were split after amplification by the preamplifier.  
This scheme enabled a simultaneous measurement of both
pulse area and time.  Resistive splitters with a frequency response flat to
within 1~dB for 0--2 GHz were used to minimize phase shifts which would
affect time measurements.  Splitting the signal resistively provided a 
6.3~dB loss in amplitude.  
The decision to provide the signals to both a TDC and an ADC was motivated 
by the need to understand the time resolution and the TDC trigger 
efficiency as a function of the number of photoelectrons present in 
the event.

A LeCroy 4413 discriminator module \cite{LeCroy} with a threshold 
setting of 
$\sim$20~mV provided discriminated signals from the VLPC channels to a 
LeCroy 2228A CAMAC TDC module with a least count of 50~ps for time 
measurements relative to the system trigger.  
A LeCroy 2249A CAMAC ADC module integrated 
the accumulated charge on each tested
channel during a gate of 80~ns after a triggered event.  


\section{Timing Test: Data Analysis and Modelling}

Analysis of the system time resolution for a specific configuration was 
straightforward given the integrated charge and pulse arrival time at each 
end of the fiber for each event. Rather than relying on the system trigger to 
provide a reference time, the time difference between signal arrival for 
each end was measured.  The system time resolution was defined as 
the root mean square (RMS)  of the distribution of time differences. 

In spite of the collimated source, most triggered events did not produce 
scintillation photons within the fiber.  
This was due to the small cross-section presented by the fiber and to 
ineffectively collimated gamma ray photons and bremsstrahlung x-rays
produced by the source.
Such events were eliminated by requiring that each end of the fiber yield a
signal greater than two standard deviations above pedestal.


Within the resolution of the instruments involved, the pulse area information 
yielded the number of photoelectrons produced in the VLPC in any given event.  
The ADC scale was determined from low photon multiplicity events where the 
pulse area distribution revealed peaks from integer numbers of photoelectrons. 
The high-bandwidth, high-gain preamplifier used for these tests provided 
relatively poor pulse area resolution due to noise in the system. 
This effect contributed to the 
systematic uncertainty assigned to the measured mean light intensity.  

\begin{figure}
\centerline{\psfig{file=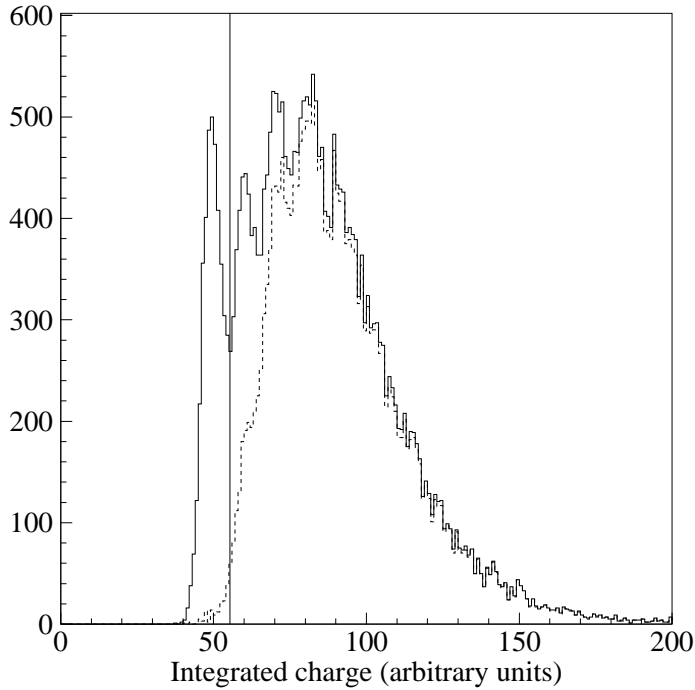,height=4.0in}}
\caption{
Integrated charge distribution for low multiplicity events.  The dotted
histogram is a subset of the solid histogram and shows events for which the 
TDC found a pulse.  The vertical line indicates two standard deviations above 
the pedestal mean.
\label{tdceff}}
\end{figure}

The presence of the ADC information also allowed a determination of the 
efficiency of the TDC discriminator to trigger on single photoelectron events. 
Figure~\ref{tdceff} is a good example of the integrated charge distribution 
for one channel for low photon multiplicity events.  The solid histogram 
represents events with a 
signal ($>$2$\sigma$ above pedestal) present on the opposite end of 
the fiber. The 
peaks in the distribution correspond to integer numbers of photoelectrons 
produced in the VLPC; the first peak is the pedestal which has not been
suppressed.  The dotted histogram is the distribution after
events which had non-physical TDC values were eliminated--- either the TDC 
fired prematurely due to noise in the circuit or did not fire at all. The
vertical line is placed at two standard deviations above pedestal; events below
this value are removed in the final analysis. With 
the chosen preamplifier, the TDC discriminator was estimated to have an 
efficiency of approximately 30\% for single photoelectron events relative 
to the efficiency for events of higher multiplicity.  

A simple Monte Carlo simulation using Mathematica~\cite{Mathematica}
was developed to predict the 
expected time difference distribution.  The model included the  light 
intensity at the VLPC, decay time of 
the scintillator and wave shifter, dispersion 
in the fiber, and random noise contributing to premature firing of the TDC's.

\begin{figure}
\centerline{\psfig{file=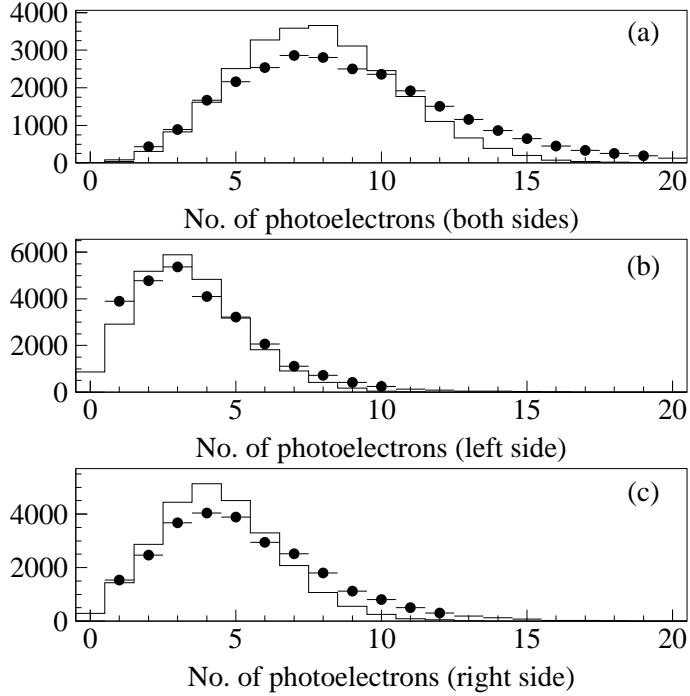,height=4.0in}}
\caption{
The number of events recorded as a function of the number of photoelectrons
detected for the longest fiber configuration. The data is depicted by the
points; the histogram represents the Monte Carlo. Plot (a)
is the combined total from both ends.  The lower plots show each end
individually. 
\label{adc_data_mc}}
\end{figure}

The light intensity for a given set of conditions was derived from the data.
The integrated charge for each fiber end was considered 
separately. 
Using the measured gain and pedestal mean, the distribution 
was scaled to yield photoelectron number rather than pulse area.  
The results
from both ends were then combined to give the total number of 
photoelectrons produced for each event.  Within the Monte Carlo, 
photon production was 
modeled for the combined total which accounted for correlations between 
light intensities at each end.  It was found
that the average light intensity at one end of a fiber was often 
different from
that of the other end.  This was assumed to be due to poor optical coupling
within one or more of the fiber connectors.  Thus within the Monte Carlo,
each photon's longitudinal direction was randomly chosen with a weighting 
determined by the average light intensities detected at each end.
For simplicity, photons were produced
according to a Poisson distribution rather than the actual probability function
given by the data.  The Poisson mean was varied by $\pm 20\%$ from the 
central value to account for this discrepancy and for the uncertainty in the 
photoelectron scale.  
Photon loss in the fiber was not simulated since this is not strongly 
correlated with the photon's time of arrival.
Figure~\ref{adc_data_mc} shows a 
comparison of the number of photons seen in the
data to the Monte Carlo model for the lowest light intensity tests.

Within the Monte Carlo simulation, photons were generated with an exponential  
time distribution in agreement with the exponential decay of the scintillator 
which has been measured to have a lifetime of $8.2 \pm 0.2$~ns.  A 
simple ray tracing algorithm was used to model the time dispersion of 
photons within the fiber.

The Monte Carlo simulation calculated the time difference between the first 
detected photons at each end of the fiber after including an overall system 
time resolution.  The measured inefficiency for 
detecting a single photon was included in the model; the discriminators were
assumed to be fully efficient for more than one photoelectron.  Studies of 
the rate of premature firing of the TDC's in the data prior to the first 
photon arrival yielded a noise probability which was also included in the 
simulation.

\begin{figure}
\centerline{\psfig{file=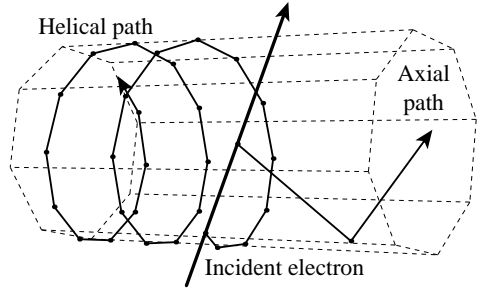,height=1.5in}}
\caption{
Isometric view of fiber showing axial and helical photon paths in
a simulated event.
The fiber is depicted as an octagonal cylinder with a greatly 
expanded diameter.  
\label{isofiber5}}
\end{figure}

\begin{figure}
\centerline{\psfig{file=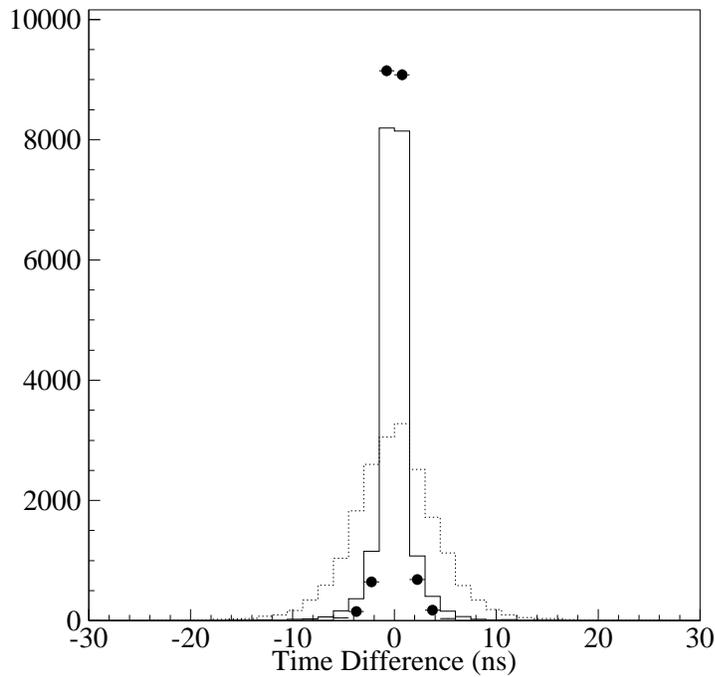,height=4.0in}}
\caption{
Monte Carlo comparison of the time difference between signal produced at 
opposite
ends of a 20.0~m scintillating fiber for axial photon 
paths (points) and all photon paths (solid histogram) assuming 
prompt photon production at the center of the fiber.  
Also shown is the distribution for events with the decay time of
the scintillator and wave-shifter included (dotted histogram).  
Exactly ten photons were produced in each event.
\label{tdc_ax_vs_hel}}
\end{figure}

Photons were assumed to be generated at the axial and longitudinal center 
of the fiber.  Allowing the photons to be generated off-axis yielded an 
approximate 20\% decrease in the average longitudinal distance 
traveled between bounces assuming a reflection probability of unity 
for photon paths within the critical angle. 
The decrease was due to helical photon paths which 
allow photons to be retained in the fiber with lower longitudinal 
velocity as shown in the Monte Carlo event
depicted in Fig.~\ref{isofiber5}.  
This effect is largest for long fibers and small photon statistics.
However, as shown in Fig.~\ref{tdc_ax_vs_hel}, the increased 
dispersion was 
not evident in the total time resolution due to the long decay time 
of the wave shifter which dominated the distribution. Including a 
reasonable absorption length and coefficient of reflection in the 
simulation further reduced the effect of the off-axis photons due to 
the higher probability that these photons are not detected.  
Thus, off-axis photons were not included since it was felt that the 
added assumptions yielded only a minor correction which likely exceeded 
the predictive power of the basic ray-tracing prescription.

\begin{figure}
\centerline{\psfig{file=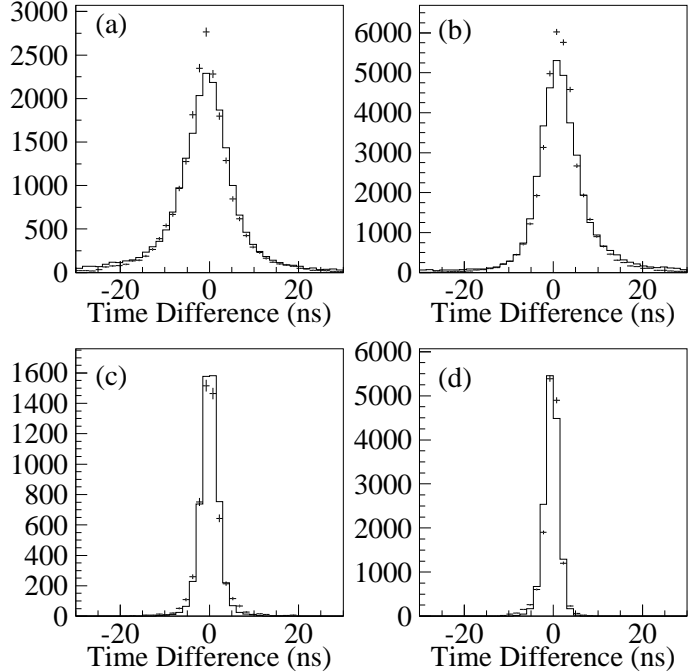,height=4.0in}}
\caption{
Time difference between signals produced by VLPC's attached to both ends 
of fiber for a total fiber distance between source and VLPC of 
(a)~9.8~m, (b)~8.8~m, (c)~1.8~m, (d)~0.8~m.
The points with error bars represent the data.
\label{tdc_mc_d}}
\end{figure}

A comparison of the the time difference spectra from data and Monte Carlo 
simulation for four different fiber lengths is shown in Fig.~\ref{tdc_mc_d}
with the points representing the data.  The Monte Carlo results incorporate 
a system time resolution of 750~ps which yielded the best overall match to the
data.  The statistical error on the Monte Carlo
prediction is suppressed but is similar in magnitude to that of the data.
The measured time resolution agrees with the prediction to 
within $\sim$25\%. 
Note that the Monte Carlo result is 
normalized to match the data; only a comparison of the shapes is meaningful.
The offset of the mean and the asymmetric shape of 
the distributions are due to the uneven distribution of light between the
two sides.  
The results of the data analysis for the four different conditions are 
summarized in Table~\ref{table_timing}.

\begin{table}
\caption{Measured time resolutions and configurations.}
\label{table_timing}
\begin{tabular}{c c c c c c}
\hline
\multicolumn{2}{c}{Fiber length from center (m)} &
\multicolumn{2}{c}{Mean Photoelectrons} & 
\multicolumn{2}{c}{Time resolution (ns)}  \\
Scintillating & Clear & Left & Right & Data & Monte Carlo \\
1.5 & 8.5 & 3.4 & 4.4 & 7.0 & 8.0\\
0.5 & 8.5 & 6.5 & 4.0 & 5.7 & 7.0\\
1.5 & 0.5 & 12.5 & 12.5 & 3.0 & 2.3\\
0.5 & 0.5 & 13.7 & 21.4 & 2.0 & 1.7\\
\hline
\end{tabular}
\end{table}

\begin{figure}
\centerline{\psfig{file=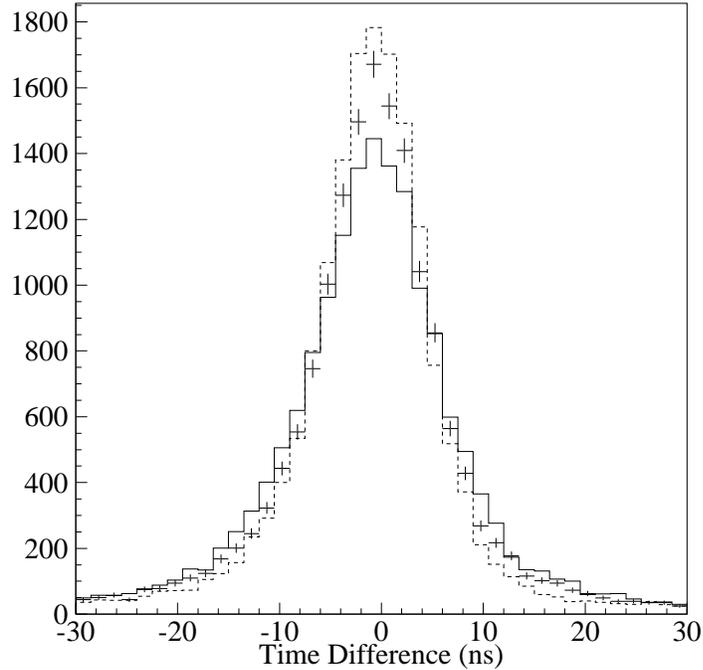,height=4.0in}}
\caption{
Time difference distribution predicted by Monte Carlo for the
longest fiber configuration.  The points represent the nominal
light intensity; the histograms give the result for a 
variation in light intensity of $\pm 20\%$.  The time resolution
prediction varies by $\pm8\%$.
\label{tdc_light_vary}}
\end{figure}

The 20\% systematic uncertainty in light intensity was not used for the 
Monte Carlo results shown  in Fig.~\ref{tdc_mc_d}. 
Figure~\ref{tdc_light_vary} shows the 
variation in time resolution for the longest fiber configuration when the 
light intensity is varied by $\pm20\%$; the nominal light intensity is given 
by the points with statistical error bars.  While both measurement
and simulation show that the time resolution depends strongly on the 
incident light intensity,  this uncertainty does not account for the 
difference between the prediction and the data.  Based on the disagreement 
observed, which can be attributed to the simplicity of the simulation, 
predictions of future performance are quoted with a systematic uncertainty 
of $\pm50\%$.

\section{Timing Test: Discussion}

The longitudinal position resolution for a charged particle impacting a
fiber detector is a linear function of the system time resolution,
\[ \sigma_{p} = K(I)\frac{c}{2n}\sigma_{t} \]
where $c$ is the speed of light and $n$ is the refractive index of the fiber.
The factor $K$ is a function of the light intensity, and in this application
represents the deviation of the average longitudinal speed of the 
first detected photon from the speed of light in the fiber, assuming an even
distribution of light. The Monte Carlo model described above
indicates that for the lowest possible light intensity with the standard fiber,
one detected photon on each end, $K \approx 0.96$.  
The value of $K$ approaches 
unity with increasing light intensity with $K \approx 0.99$ for ten total
photoelectrons.  The following discussion assumes $K = 1$.

Tests with cosmic ray muons indicate that a minimum ionizing particle 
typically produces $\sim$10 total photoelectrons for the longest 
fiber configuration tested here.
The results given
above then indicate a position resolution of $\sim$60~cm for a minimum 
ionizing particle.  Thus time measurements
are not a reasonable option for longitudinal position determination in
the type of detector constructed for these tests.

Comparison of the simulation with the data suggests that the dominant effects
determining the time resolution of the system are the light intensity
and the decay time of the scintillator.  
Time dispersion of photons within the fiber is of secondary importance.
Thus, dramatic improvements would
arise from a faster wave-shifter or more light production.
Furthermore, 
a new version of the VLPC, HISTE-V, has a factor of $\sim$2.5 higher 
gain which should provide 100\% efficiency for time measurements with
a single photoelectron; this should effect a modest improvement in time
resolution for low multiplicity events.  

\begin{figure}
\centerline{\psfig{file=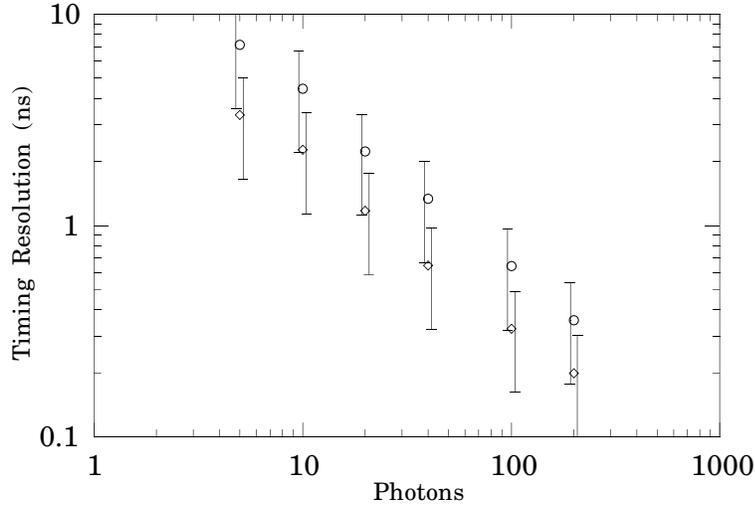,height=2.8in}}
\caption{
Monte Carlo prediction of expected time resolution for future systems.
The circles are for the standard fiber described above.  The diamonds are 
for scintillating fiber with a faster fluorescence decay time of 2.7 ns.
The error bars are offset for clarity and
denote the assigned 50\% systematic error.
\label{mc_tim_res}}
\end{figure}

The preceding observations motivated a final set of simulations yielding 
predictions of the time resolution as a function of total 
number of photoelectrons produced for two detector configurations: the first 
using the 3HF fiber, the second using commercially available
green scintillating fiber with a decay time of 2.7~ns.  Unfortunately, 
the fast scintillating fibers that are available have 
attenuation lengths that are small compared to the 3HF fiber.  
The results are shown in Fig.~\ref{mc_tim_res}.
These results follow from the ideal case where the electronics time resolution
is negligible and the TDC's are assumed to 
be 100\% efficient for detecting a single photoelectron.  The photons were 
assumed to be produced in the longitudinal center of a
fiber of total length 20~m.


These final results provide information necessary to understand the 
effectiveness of using time difference measurements to determine the
longitudinal position of a particle or shower in a scintillator 
detector.  As an example, the central fiber tracker proposed 
for the \D0 upgrade will be constructed of scintillating fiber identical
to the standard fiber described in Table~\ref{table_fiber}.  
For a single fiber, 
the expected ten photoelectrons produced by a minimum ionizing particle yield 
an RMS longitudinal position resolution of $\sim$50~cm.    
\D0's proposed tracking detector would eliminate gaps by 
densely packing two layers of fibers for each superlayer. 
Instrumenting fiber doublets would then increase the mean number of 
photoelectrons to 20 and should improve the position resolution to 
$\sim$27~cm for each superlayer.  This resolution could be useful to 
eliminate combinatoric background in a tracking-based trigger. 
As another example, the central preshower detector proposed for the \D0 upgrade
is expected to have a minimum trigger threshold of $\sim$80~photoelectrons
for electron candidates~\cite{Lincoln}.
Using wavelength shifting fiber with the same fluorescence decay time 
as that used for the tracker would yield an 
RMS longitudinal position resolution of $\sim$12~cm. 
This resolution is acceptable for certain trigger requirements and, 
like all of the scenarios discussed here,  could be dramatically
improved by using fiber with a faster decay time.  

\section{Gain Linearity Test: Experimental Setup}

\begin{figure}
\centerline{\psfig{file=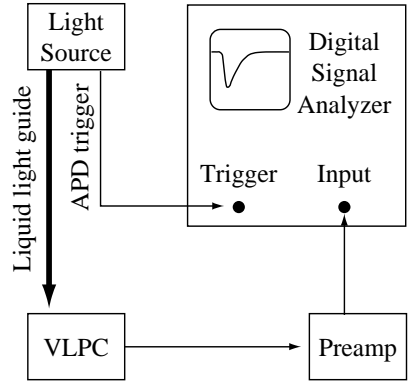,height=2.0in}}
\caption{
Block diagram of experimental setup for VLPC gain linearity studies.
\label{linblock}}
\end{figure}

The linearity tests shared the VLPC cassette and cryogenic system with the 
timing tests; other details of the system differed.  A block diagram of the 
experimental setup is shown in Fig.~\ref{linblock}.

\begin{figure}
\centerline{\psfig{file=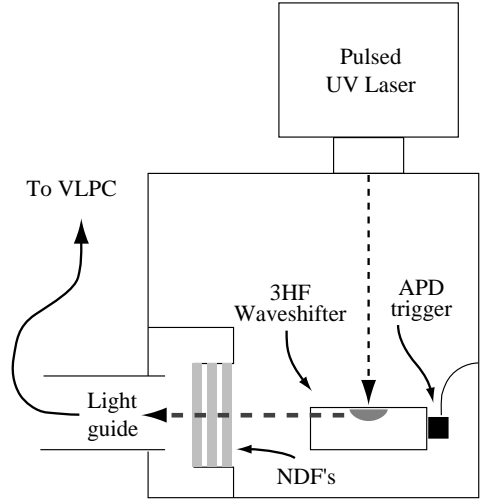,height=2.6in}}
\caption{
Block diagram of the light source used for VLPC gain linearity studies.
\label{linset4}}
\end{figure}

Because timing was not an issue for this study, bulk scintillator with 
the same composition as the fiber core was used as the light source.  
A schematic of the light source is shown in Fig.~\ref{linset4}. The 
scintillator was illuminated with an ultraviolet (UV) nitrogen laser 
producing pulsed light ($\lambda$~=~337.1~nm) with a width of 3~ns 
at a frequency of 10~Hz. The UV light activated the wavelength shifting 
component of the scintillator which subsequently produced visible 
light. This configuration provided a constant intensity light source 
with the appropriate wavelength spectrum. The low pulse frequency 
ensured adequate recovery time for the VLPC. Absorptive neutral 
density filters (NDF's) attenuated the light by known factors and served 
as the reference by which linearity was judged.  A liquid light guide 
transported the light from the source to the VLPC cassette.  

\begin{figure}
\centerline{\psfig{file=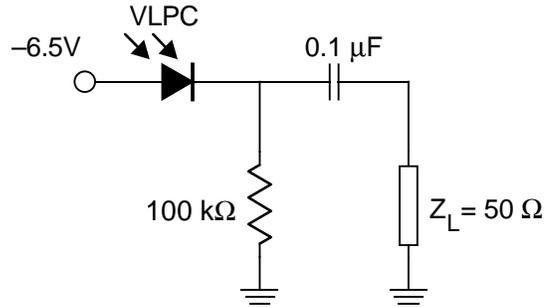,height=1.6in}}
\caption{
Schematic of the circuit used to connect the digital signal analyzer directly
to the VLPC for the gain linearity studies.
\label{vlpc_dir}}
\end{figure}

Three preamplifier configurations were used to ensure that the 
limited dynamic range of any single amplifier did not affect the measurement.
The highest gain preamplifier was similar to that used for the timing tests 
but lacked the second stage of amplification; this improved its dynamic
range.   The mid-gain preamplifier was of the same 
design but attenuated the signal from the VLPC at the input.  The lowest gain 
``preamp'' was a circuit allowing the VLPC to be directly read by
the input stage of the digital signal analyzer.
Thus, the signal was not amplified at all, and any saturation 
seen at this point was due to the VLPC rather 
than any intervening electronics. 
Figure~\ref{vlpc_dir} shows a schematic of the low gain circuit.

A commercial digital signal analyzer (DSA), 
Tektronix model DSA~602A~\cite{Tektronix},
was used to integrate the charge from the output stage of the preamplifier.
The signal was averaged over the 512 events prior to the measurement with the 
systematic uncertainty being determined
by varying the gate width from 40~ns to 80~ns.
The measurement error of the DSA was small compared to the 
systematic uncertainty assigned to the measurement.

The photoelectron scale was set in the same manner as 
described above for the timing
measurements using a CAMAC ADC to record the integrated charge 
distribution for the lowest intensity configuration.  The mean number of
photoelectrons from this distribution then served as 
the reference for all configurations since the filters attenuated 
the signal by a known
amount.  A 10\% uncertainty was assigned to the optical density of the 
filters in accordance with the manufacturer's specifications.
The measurement of the mean number of photoelectrons for the 
reference configuration was assigned a systematic uncertainty of 
15\% which is fully correlated for all points.

\section{Linearity Test: Results and Discussion}

\begin{figure}
\centerline{\psfig{file=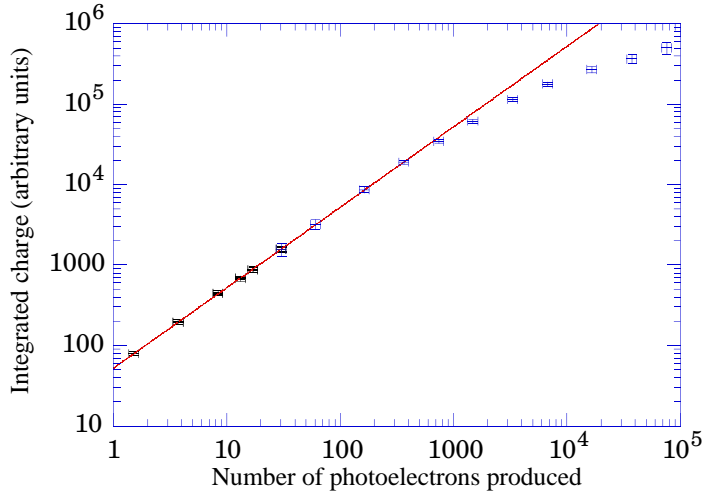,height=2.6in}}
\caption{
VLPC response vs. input photoelectrons.
\label{newch114lin}}
\end{figure}

\begin{figure}
\centerline{\psfig{file=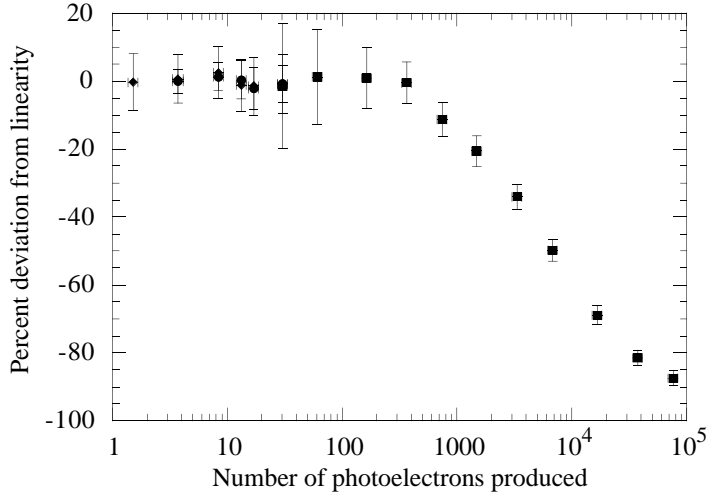,height=2.6in}}
\caption{
Percent deviation from linearity for VLPC vs. input photoelectrons.
Data from all three tested preamp configurations are shown.
\label{newch114percdev}}
\end{figure}

Two channels contributed to the linearity measurement.
Figure~\ref{newch114lin} shows the response of the VLPC as a function of the
number of 
photoelectrons produced for one channel.  Figure~\ref{newch114percdev}
shows the percent deviation from linearity for the same channel.
The uncorrelated uncertainties
are given by the error bars; there is an additional 15\% uncertainty in the
photoelectron scale.  This VLPC is linear to within 10\%
out to $\sim$600 
photoelectrons. A similar result was obtained from the 
other tested channel. 

The VLPC provides high gain due to fast avalanche 
multiplication of charge within the device.  
As described in the literature~\cite{Stapelbroek1,Stapelbroek2}, the avalanche
produces a space charge in a small area inhibiting subsequent 
avalanches within the recovery time of the device which could be as long as 
1~ms.  
A Monte Carlo model was developed to characterize the surface size of two 
idealized regions: a small region for which subsequent avalanches were fully 
inhibited yielding a dead area, and a larger region for which the average 
gain for subsequent avalanches was reduced by 10\%.  The model assumed that 
the affected regions were circular, and that the recovery time of the device 
was long compared to the time distribution of photons in a given event.  
To characterize the size of the smaller region, it was assumed that the
avalance due to a single photoelectron left the circular area completely dead
for subsequent photons.  
In such a model, complete coverage of the active area of the VLPC
would result in saturation.  Assuming that full saturation begins for
$\sim$1$\times 10^{5}$ photoelectrons as suggested by 
Fig.~\ref{newch114lin}, the simulation reveals the 
radius of the dead region to
be 1~$\mu$m.
This is consistent with theoretical calculations~\cite{Stapelbroek3}.
The small size of the dead region also suggests an 
explanation for the deviation
from linearity that is observed to begin at $\sim$600 photoelectrons.
Since the gain region is only 5~$\mu$m 
thick, the space charge is concentrated in a region that is only 
1--2~$\mu$m on a side.  This clump of positive charge will then distort the 
local field causing some loss of amplification for nearby photoelectrons that
are outside the dead region.
For an average gain reduction of 10\%, 
the same Monte Carlo model was used to estimate the diameter 
of the affected circular region at 23~$\mu$m.
This result must
be interpreted as the average cut-off diameter since the model assumes 
a rigid cut-off whereas the field distortion will likely cause a 
continuous reduction in gain.  The results given by this model allow a 
prediction of VLPC linearity for systems with different coverage 
of VLPC pixels than the apparatus described here.

These results indicate that the VLPC is an ideal 
device for calorimetric applications. It can provide high
sensitivity and can accept large incident light intensities.

\section{Conclusion}

The time resolution was measured for a scintillating fiber detector
instrumented on both ends with HISTE-IV VLPC's operated at 6.5 K and 
with a bias voltage of 6.5 V.  The measurements were shown to agree 
with predictions from a simple Monte Carlo simulation incorporating only 
the decay time of the scintillator, the mean light intensity, 
and the dispersion in the fiber as modeled with a simple ray-tracing
algorithm.  The light intensity coupled with the 
decay time of the scintillator
clearly dominates the resolution.
The results of the study suggest that time difference 
measurements are not attractive for longitudinal position measurements
in single fibers unless the fiber has a short fluorescence decay time 
($\lesssim$2~ns).  Higher light intensities (which are expected in 
electromagnetic showers) provide a more acceptable position resolution.

Measurements also addressed the linearity of HISTE-IV VLPC's over a wide 
range of incident light intensities.  With a bias voltage of 6.5 V at 6.5 K, 
the VLPC was shown to have a linear response up to
$\sim$600 photoelectrons, and the VLPC output continued to increase
monotonically up to incident light intensities producing
at least $7 \times 10^{4}$ photoelectrons.  

\section{Acknowledgements}
Technical assistance for this project
was provided by the \D0 department of the Fermilab Research Division.
The authors acknowledge the support provided by the U.S. 
Department of Energy, the University of Maryland, and CNPq in Brazil.


\begin{thebibliography}{99}
\bibitem{model}
The measurements reported are valid only for the Rockwell HISTE-IV model VLPC 
operating under the conditions specified.
\end{thebibliography}
\end{document}